\documentclass[aps,prl,amsmath,amssymb,superscriptaddress,showpacs,10pt,reprint]{revtex4-1}

\usepackage{multirow}

\usepackage[dvips]{graphicx}
\usepackage[colorlinks=true]{hyperref}
\usepackage[normalem]{ulem}

\graphicspath{{Figures/}}
\hypersetup{pdfpagemode = UseNone}

\newcommand{\setup}[1]{{\textsf{#1}}}
\newcommand{\eg}{e.g.,~}
\newcommand{\ie}{i.e.,~}

\newcommand{\Tc}{T_\mathrm{c}}

\newcommand{\teff}{T}
\newcommand{\aPhi}{\Phi}

\newcommand{\ket}[1]{\left|#1\right\rangle}

\begin{document}
\title{Measurement of a microwave field amplitude beyond the  standard quantum limit.}

\author{M.~Penasa}
\author{S.~Gerlich}
\author{T.~Rybarczyk}
\author{V.~M\'etillon}
\author{M.~Brune}
\author{J.M.~Raimond}
\author{S.~Haroche}
\affiliation{Laboratoire Kastler Brossel, Coll\`ege de France, CNRS, ENS-PSL Research University, UPMC-Sorbonne Universit\'{e}s, 11, place Marcelin Berthelot, 75231 Paris Cedex 05, France}
\author{L.~Davidovich}
\affiliation{Instituto de F\'{i}sica, Universidade Federal do Rio de Janeiro, 21941-972, Rio de Janeiro (RJ), Brazil}
\author{I.~Dotsenko}
\affiliation{Laboratoire Kastler Brossel, Coll\`ege de France, CNRS, ENS-PSL Research University, UPMC-Sorbonne Universit\'{e}s, 11, place Marcelin Berthelot, 75231 Paris Cedex 05, France}
\email{igor.dotsenko@lkb.ens.fr}
\date{\today}

\begin{abstract}

We report a quantum measurement beyond the standard quantum limit (SQL) for the amplitude of a small displacement acting on a cavity field. This measurement uses as resource an entangled mesoscopic state, prepared by the resonant interaction of a circular Rydberg atom with a field stored in a superconducting cavity.  We analyse the measurement process in terms of Fisher information and prove that it is, in principle, optimal. The achieved experimental precision, $2.4$~dB below the SQL, is well understood in terms of experimental imperfections.
This method could be transposed to other systems, particularly to circuit QED, for the precise measurements of weak forces acting on oscillators. 

\end{abstract}

\maketitle

\section{I. Introduction}

Metrological measurements are of paramount importance in fundamental physics and technology. They generally rely on the estimation of the value of a parameter~$\beta$ (\eg the amplitude of a dc or ac electromagnetic field, a small mechanical force, \textit{etc}) controlling the evolution of a quantum system. This system, initially prepared in a `resource' state, evolves according to the parameter value and is finally measured, directly or indirectly through ancillae.

Due to the intrinsically statistical nature of the quantum measurement, the final standard deviation, $\Delta\beta$, of the parameter estimation scales as $1/\sqrt \nu$, in the limit of a large number $\nu$ of experimental realizations: $\Delta\beta=\Delta\beta^{(1)}/\sqrt \nu$ (saturated Cram\'{e}r--Rao bound \citep{Cramer,Rao}). Here, $\Delta\beta^{(1)}= 1/\sqrt F$, where $F$ is the Fisher information provided by a single realization of the measurement protocol. 

For a given resource state, $F$ is bounded from above by the quantum Fisher information, $F_Q$. It measures the maximal information on $\beta$ which can be imprinted onto the resource state and is independent upon the final measurement procedure (quantum Cram\'{e}r--Rao bound \citep{Braunstein94}). Optimizing the measurement precision amounts to choosing the resource state so that $F_Q$ is large and to choosing the final system's measurement to realize $F~\!=~\!F_Q$.

When the resource state is classical (\eg a coherent state for a harmonic oscillator), $F_Q$ defines the standard quantum limit (SQL) \cite{Jaekel90}. This limit can be overcome by using a non-classical resource state \cite{Giovannetti04}, such as a squeeezed state \cite{squeezed} or a mesoscopic quantum state superposition (MQSS) \cite{Facon16}. This strategy has led to a considerable development for  quantum-enabled metrology~\cite{Giovannetti11} beyond the SQL. Among the many remarkable achievements of this active field, let us mention sensitive optical phase measurements \cite{phase}, magnetometry \cite{magnetometry}, 
and gravitational wave detection \cite{GravWaves}.

A particularly interesting class of measurements is that of a weak force acting on an oscillator-like system and resulting in a small displacement of the resource state \cite{Caves80,Latune13}. It is relevant for the detection of small forces in the optomechanical context \cite{Aspelmeyer14}, of a photon scattering recoil in ion traps \cite{Hempel13}, and of weak fields in spin systems~\cite{Sewell12}.

For harmonic oscillator displacements, the SQL is simply determined by the extension of the Wigner distribution in phase space of a classical coherent state, of the order of $\sqrt\hbar$.~As shown in Ref.~\cite{Zurek01}, beating the SQL thus amounts to using a resource state whose Wigner representation has structures at a scale lower than $\sqrt\hbar$, \ie sub-Planck structures, conspicuous in squeezed states or in MQSS.  

In this paper, we report the first quantum-enabled  measurement of a microwave field amplitude based on a mesoscopic non-classical resource state of an entangled atom-cavity system. It uses the resonant interaction between an initially coherent field in a superconducting cavity and a single circular Rydberg atom, as proposed in Ref.~\cite{Toscano06}. This interaction prepares a MQSS entangled state, which is used as a resource for measuring the amplitude of a microwave field injected into the cavity, and leads to a quantum Fisher information much larger than that resulting from the initial coherent state.
The resource state undergoes the displacement by an amplitude $\beta$ to be measured. The subsequent atom-field interaction and the final state-selective atomic detection lead to a quantum measurement approaching the quantum Cram\'{e}r--Rao bound. The precision $\Delta\beta^{(1)}$ is found to beat the SQL, by 2.4 dB. This quantum-enabled measurement protocol could be fruitfully transposed in other contexts, particularly that of circuit QED.

The paper is organized in the following way. Section~II describes in more details the measurement protocol. Section III analyzes the measurement in terms of Fisher information and shows that it ideally saturates the quantum Cram\'{e}r--Rao bound. Section IV is devoted to the description of the experiment and Section V to a discussion of its results. We finally conclude in Section~VI.

\section{II. Measurement protocol}

The aim of this experiment is to measure the amplitude $\beta$ of a small displacement produced by a classical source coupled to a cavity. Along the lines of Ref.~\cite{Toscano06}, we use as measuring system a two-level atom (upper state $\ket e$, lower state $\ket g$) and a field stored in the cavity. The resource state is produced by the resonant interaction during a time $T_1$ of the atom, initially in $\ket e$, with a coherent field $|\alpha\rangle = e^{-\alpha^2/2} \sum_n (\alpha^n/\sqrt{n!}) |n\rangle$, where $|n\rangle$ is the Fock state with $n$ photons and $\alpha$ is taken as real without loss of generality. 

The atom undergoes in the initial coherent field a quantum Rabi oscillation entangling it with the cavity. In an approximation valid for a large enough $\alpha$ and for moderate interaction times, the atom-field state $|\Psi\rangle $ after interaction time $T_1$ reads
    \begin{equation}\label{eq:entanglement}
         |\Psi\rangle \simeq \frac{1}{\sqrt{2}}\left[e^{-i\aPhi\alpha^2}|\alpha^+\rangle |\Psi^+\rangle 
											   -e^{i\aPhi\alpha^2} |\alpha^-\rangle |\Psi^-\rangle \right],
    \end{equation}
where $\aPhi = \Omega_0T_1/4\alpha$ and where the field and atomic states are 
    \begin{eqnarray}\label{eq:evolFirstStates}
        |\alpha^\pm\rangle  &=& |\alpha e^{\mp i\,\aPhi}\rangle,\\
        |\Psi^\pm\rangle    &=& \frac{1}{\sqrt{2}}\left[e^{\mp i\,\aPhi}|e\rangle\pm|g\rangle\right],
    \end{eqnarray}
respectively \cite{Auffeves03}. The field is thus split into two coherent components, $|\alpha^\pm\rangle$, which rotate in opposite directions in phase space. 

For small values of $T_1$, these two coherent fields still partially overlap. The atom and the cavity are not yet maximally entangled and the population of state $\ket g$ undergoes a Rabi oscillation at the average frequency $\Omega_0\sqrt{\alpha^2+1}$. As the two components separate further, the atom-cavity entanglement grows and the Rabi oscillations accordingly collapse after the characteristic collapse time $T_c = 2\sqrt{2}/\Omega_0$. For $T_1>T_c$, the two field components are nearly orthogonal, and the atom-field system is cast in a MQSS.

\begin{figure}[t]
    		\includegraphics[width=6cm]{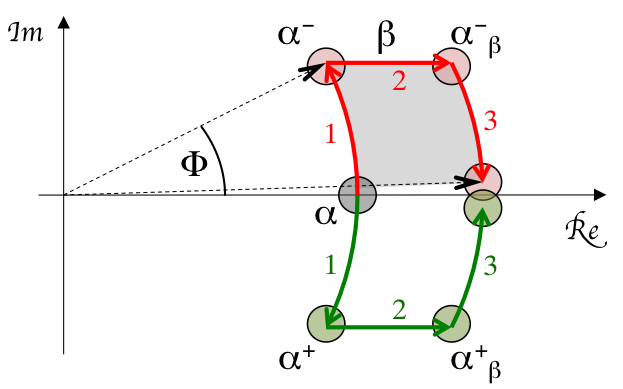}
  		\caption{Schematic evolution of the cavity field in phase space during the experimental protocol measuring small coherent field displacement $\beta$. See text for details.}
  		\label{fig:phasespace}
	\end{figure}

Figure~\ref{fig:phasespace} schematically shows the evolution of the field in phase space starting from the initial state $|\alpha\rangle$. The creation of the resource MQSS corresponds to two arrows labelled $1$. After time $T_1$, we perform the displacement by a real amplitude $\beta$, both field components being changed into  $\vert{\alpha^\pm_\beta}\rangle=e^{-i\alpha\beta\sin\Phi}\ket{\alpha^\pm+\beta}$, see arrows labelled $2$.

The measurement of the system starts after this injection. It relies on the observation of a revival of the Rabi oscillation. As shown  in Refs.~\citep{Morigi02, Meunier05}, the Rabi signal can be revived after its initial collapse by applying a time inversion, induced by  a $\pi$-phase shift between atomic states $\ket g$ and $\ket e$. This inversion results in an atom-cavity state
   \begin{equation}\label{eq:entanglement2}
         |\Psi\rangle = \frac{1}{\sqrt{2}}\left[e^{-i\aPhi\alpha^2}|\alpha^+_\beta\rangle |\Psi'^-\rangle 
											   -e^{i\aPhi\alpha^2} |\alpha^-_\beta\rangle |\Psi'^+\rangle \right],
    \end{equation}
with the new atomic states reading
\begin{equation}
|\Psi'^\pm\rangle    = \frac{1}{\sqrt{2}}\left[e^{\mp i\,\aPhi}|e\rangle\mp|g\rangle\right].
\end{equation}

Due to this atomic phase flip, the subsequent field evolution is time-reversed from that during time $T_1$. The two components of the field MQSS  merge again  (arrows $3$ in Fig.~\ref{fig:phasespace}) for a measurement time $T_2$ around  $T_1$. At the end of this period, the interaction is stopped by detuning the atomic frequency out of the cavity resonance and the atomic state is measured in the $\{\ket g, \ket e\}$ basis. 

For $T_2=T_1$, the final probability, $P_g$, for finding the atom in state $\ket g$ has the following simple expression:
    \begin{eqnarray}\label{eq:PgSymmetric}
        P_g &=& \frac{1}{2}\Big\{1+\cos(2D\beta)\Big\}\approx\frac{1}{2}\Big\{1+\cos(\Omega_0T_1\beta)\Big\}\ ,
    \end{eqnarray}
where $D=2\alpha\sin\aPhi$ is the separation, in phase space, of the field components $\vert\alpha^\pm\rangle$ before the measurement. These expressions hold when $D$ is notably larger than~$1$ (atom-field entanglement condition)  and when $\aPhi$ is not too large (implying that $\alpha$ is large) so that $D\approx \Omega_0T_1/2$. The $P_g$ signal is an oscillatory function of $\beta$, providing direct information on the displacement amplitude. Note that for large initial field amplitude $\alpha$, the oscillation period is independent of $\alpha$.

It is noteworthy that the oscillation phase, $2D\beta$, is about 4 times the shaded area in Fig.~\ref{fig:phasespace}, and reflects the geometric phase accumulated by the MQSS coherent components during their excursion in phase space. Clearly, this area, and hence the sensitivity, are maximal when the phase of the initial coherent state matches that of the measured displacement.

Moreover, these oscillations are not limited to small values of $\beta$, allowing one in principle to measure arbitrarily large field amplitudes with the same high precision. Note that the $P_g$ oscillation period, $\pi/D$, is the same as that of the oscillations observed close to the origin in phase space for the Wigner function of the MQSS $(|\alpha^+\rangle + |\alpha^-\rangle)/\sqrt{2}$. Indeed, such states known as `photonic cat' states \cite{Deleglise08} can also be used for sub-Planck metrology. However, the corresponding methods are limited to small $\beta$ values, $\beta<1$.

In the general case of $T_2\not = T_1$, the final probability $P_g$ reads
 \begin{eqnarray}\label{eq:PgAsymmetric}
        P_g &=& \frac{1}{2}\Big\{1 + C \cos(\gamma)\Big\},
    \end{eqnarray}
where 
 \begin{eqnarray}
\gamma &=&  \Omega_0 T_2 \beta + \Omega_0 \alpha (T_2-T_1)\ .
    \end{eqnarray}
The  contrast $C$ of this oscillating function of $\beta$ is set by the overlap of the coherent field components at $T_2$, given~by
    \begin{equation}\label{eq:C}
        C = \exp\big\{-\Omega_0^2(T_1-T_2)^2/8\big\}.
    \end{equation}
The highest sensitivity for a given resource state (fixed $T_1$ and large $\alpha$) is obtained by compromising a faster oscillation frequency (obtained for large $T_2$ values) and the decay of $C$ for $T_2>T_1$.

\section{III. Fisher information }

Let us now discuss the performance of this measurement in terms of Fisher information.  The absolute quantum limit for a given resource state is set by the quantum Fisher information, $F_Q$. Following Ref.~\cite{Davidovich11,Davidovich11a}, the QFI is linked to the variance in the resource state of the operator $\hat h=-i(\hat a^\dagger-\hat a)$ generating the unitary displacement $\hat D(\beta) = e^{\beta (\hat a^\dag - \hat a)}$:
    \begin{eqnarray}\label{eq:QuantumFisher}
        F_Q &=& 4 \langle (\Delta\hat{h})^2 \rangle\ .
    \end{eqnarray}
Using the resource state of Eq.~\eqref{eq:entanglement}, and for the same approximation as above, we obtain
    \begin{eqnarray}\label{eq:QuantumFisher2}
        F_Q &=& 4 (1 + D^2)\ .
    \end{eqnarray}
    
The smallest value of $F_Q$ is 4, corresponding to a coherent resource state ($T_1=0$). It thus defines the SQL and leads to $\Delta\beta^{(1)}_\textrm{SQL}=0.5$. Increasing the resource size $D$ and using a proper measurement, we go beyond the SQL and enter the sub-Planck region.  Ultimately, for $\aPhi=\pi/2$ reached at $T_1=2\pi\alpha/\Omega_0$, the size $D$ is maximal ($D=2\alpha$) and $F_Q\approx16\alpha^2$. This corresponds to the Heisenberg limit (HL) in this context. Since in our experiment, as will be discussed later, we are technically limited by the interaction duration, rather than by the resource energy, from now on, we focus  on moderately large values of $T_1$ so that
    \begin{eqnarray}
        F_Q &\approx& 4 (1 + \Omega_0^2T_1^2)\ .
		\label{eq:QuantumFisher3}
    \end{eqnarray}
    
The actual information extracted by the measurement protocol is measured by the Fisher information (FI) of the atomic signal. For a discrete measurement with two possible outcomes $s\in\{g,e\}$, this FI is given by
    \begin{eqnarray}\label{eq:FisherData}
        F(\beta) &=& \sum_s P_s(\beta)\left(\frac{\partial}{\partial \beta} \ln P_s(\beta)\right)^2.
    \end{eqnarray}
Using \eqref{eq:PgAsymmetric}, we get
    \begin{equation}\label{eq:FisherFullyAnalytical}
        F(\beta,T_1,T_2) = C^2 \Omega_0^2 T_2^2\, \frac{\sin^2(\gamma)}{1 - C^2\cos^2(\gamma)}.
    \end{equation}
    
The variation of $F$ with $\gamma$, except for $C=1$, reflects the oscillations of $P_g$. Maximum information is obtained at the `mid-fringe' points where $P_g=1/2$, \ie $\cos\gamma=0$ or $\gamma=\pi/2+p\pi$ with $p$ integer, and we get then
    \begin{equation}\label{eq:FisherFullyAnalytical2}
        F_\textrm{max} (T_1,T_2) =C(T_1,T_2)^2\Omega_0^2 T_2^2.
    \end{equation} 
It is easy to show that, as expected, $ F_\textrm{max} (T_1,T_2)$ is always lower than $F_Q(T_1)$. Getting the maximum information results from a compromise between two opposite trends. On the one hand, the quantum phase accumulated on the coherent components trajectories increases linearly with $T_2$. On the other hand, the contrast $C$ decreases rapidly when $T_2$ increases above $T_1$. The maximum resulting from this compromise indeed approaches the $F_Q$ limit for large enough $D$ values (the difference is below $1.8\%$ for $D>2$).


\section{IV. Experimental set-up}

    \begin{figure}[t]
    		\includegraphics[width=\columnwidth]{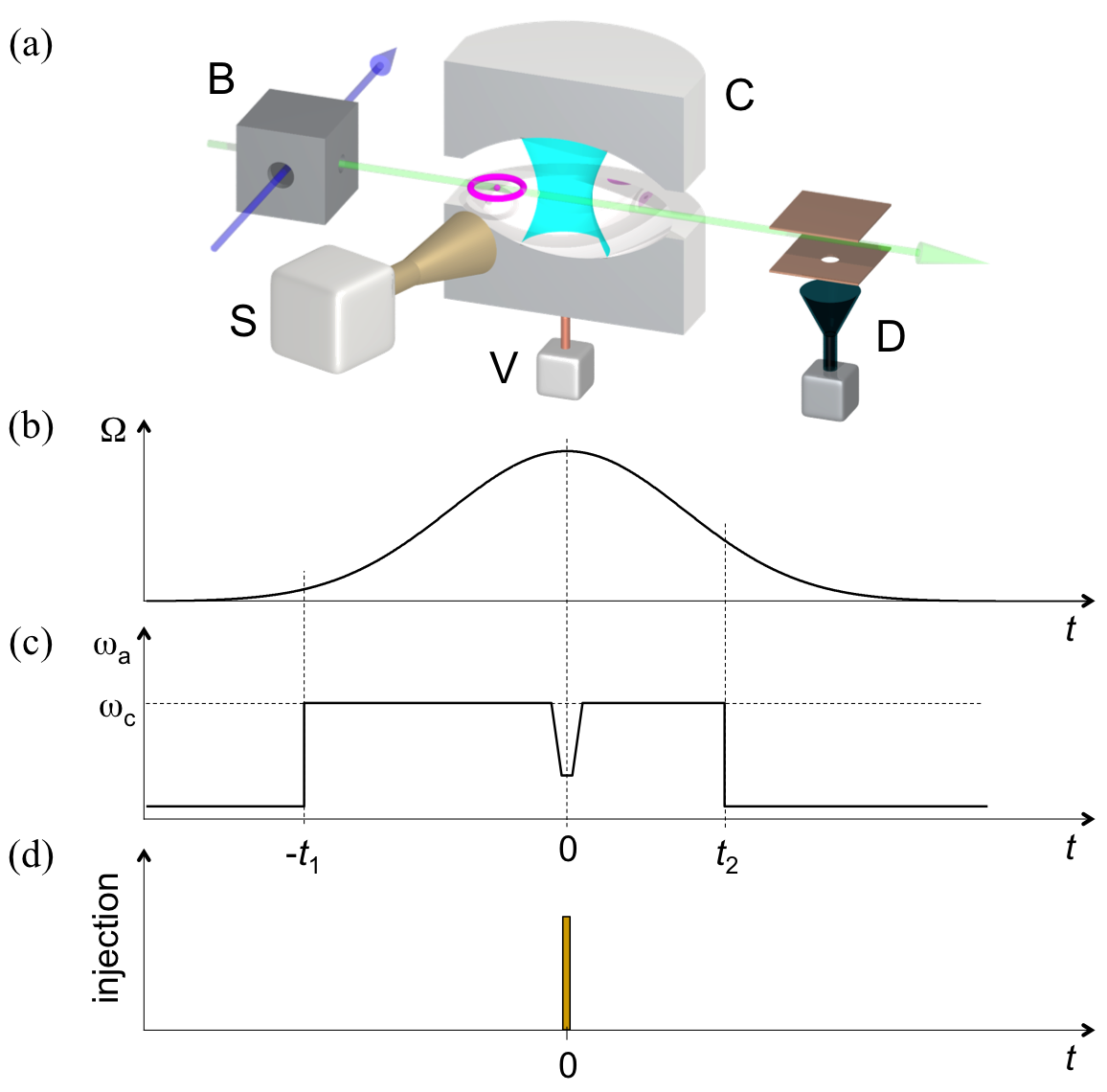}
  		\caption{(a) Scheme of the experimental setup. Coherent field is injected into a high-$Q$ cavity \setup{C} by a microwave source \setup{S}. A single Rydberg atom (magenta toroid) flying across the cavity mode is prepared from atomic beam in \setup{B} and its state is detected in \setup{D}. The atomic frequency $\omega_a$ is tuned via Stark shifting atomic levels in electric field applied between the cavity mirrors by voltage source \setup{V}. (b) and (c) Temporal variation of the atom-cavity coupling $\Omega$ and modulation of the atomic frequency $\omega_a$, respectively. (d) Coherent field $\beta$ to be measured is injected in \setup{C} at time $t=0$.}
  		\label{fig:setup}
	\end{figure} 
	
The scheme of the experimental setup is presented in Fig.~\ref{fig:setup}(a). The field is stored in a  high-$Q$ superconducting cavity \setup{C}. Its resonant frequency is $\omega_c/2\pi = 51.1$~GHz and its energy damping time is $\Tc = 65$~ms. The cavity is cooled down to 0.8 K with 0.06 thermal photons per mode on the average. The injection into \setup{C} is made by the classical microwave source \setup{S} via diffraction on cavity mirrors' edges.

The levels $|g\rangle$ and $|e\rangle$  are the circular Rydberg levels with principal quantum numbers 50 and 51, respectively. The $|g\rangle\rightarrow |e\rangle$ transition is resonant with  \setup{C}. The atom is initially prepared in  $|g\rangle$ in box \setup{B} from a thermal beam of ground state Rubidium atoms. After having interacted with \setup{C}, the atomic states are selectively detected by field ionization in ~\setup{D}.

The cavity Gaussian mode has a waist $w=5.96$~mm. The atom-cavity vacuum Rabi frequency at the cavity center is $\Omega_0/2\pi = 46$~kHz. The atom crosses \setup{C} with a $v = 250$~m/s velocity. The temporal variation of the  atom-cavity coupling is thus $\Omega(t)=\Omega_0\exp\big[ -v^2t^2/w^2 \big]$, where the time origin is set when the atom is at the cavity center, see Fig.~\ref{fig:setup}(b). It is convenient to define an effective interaction time, $\teff$. Between the times $t$ and $t'$, it is given by $\teff(t,t')= \int_{t}^{t'} \exp\{-\left(v\tau/w\right)^2\}d\tau$. The maximal interaction time corresponding to the whole cavity mode extension is thus $\teff_\textrm{max}=\sqrt{\pi}\,w/v \approx 42\,\mu$s. From now on all interaction times are given in terms of effective times.

The atomic resonance frequency, $\omega_a$, is controlled via the Stark shift produced by an electric potential difference  \setup{V} applied across the mirrors. This allows us to quickly switch on and off the resonant atom-cavity interaction and thus to control its duration, as shown in Fig.~\ref{fig:setup}(c). Besides, the same control is used to realize the atomic phase shift operation around time $t=0$.

    \begin{figure}[t]
    		\includegraphics[width=\columnwidth]{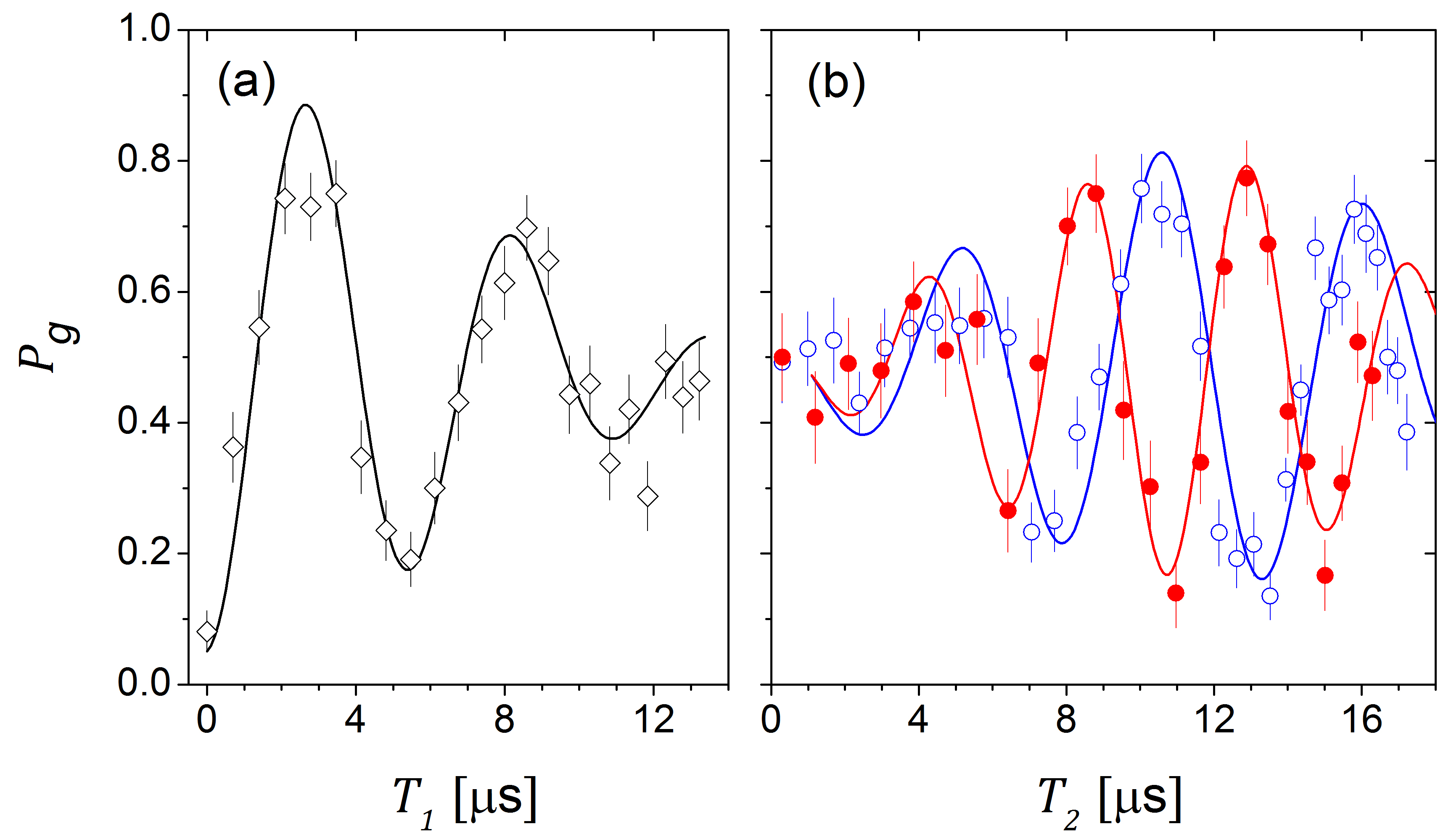}
  		\caption{Quantum Rabi signals. (a) Collapse of the $P_g(T_1)$ oscillations. Diamonds: experimental values (error bars are statistical). The solid lines result from a numerical integration of the atom-field evolution with no free parameters. (b) Revival of the Rabi oscillations $P_g(T_2)$ for $T_1=13.4\,\mu$s. Open blue circles are experimental with statistical error bars. Full red circles: induced revival of the Rabi oscillation after injecting a small coherent field $|\beta=1\rangle$ into \setup{C}  at $T_2=0$. Solid lines as in (a).}
  		\label{fig:revival}
	\end{figure}
			
We first investigate the collapse of the Rabi oscillations. We initially inject in \setup{C} a coherent field $|\alpha\rangle$ with $12.7$ photons on the average, $\alpha = \sqrt{12.7}$, and then send an atom in $|g\rangle$. We let the atom and the cavity interact for a time $T_1$ and record $P_g(T_1)$ by repeating the experimental sequence $1000$ times for each $T_1$ value. Figure~\ref{fig:revival}(a) shows the evolution of $P_g(T_1)$ (open diamonds). It exhibits the collapse due to the photon number dispersion in $\alpha$. The solid line is the result of a numerical integration of the atom-field interaction taking into account the limited state resolution of the detection (wrong state attribution in $5\%$ of cases) and the longitudinal spread of the atomic sample  (about 1~mm), which is non negligible at the scale of the cavity mode waist.

We now proceed with the complete sequence, involving the revival of the Rabi oscillations induced by the atomic phase flip. The atom, initially in $\vert g\rangle$, interacts first resonantly with the cavity for time $T_1$. When it reaches the cavity centre, at time $t=0$, a short voltage pulse is applied across the mirrors. It detunes the atomic frequency by $1.25$~MHz during a $0.4\,\mu$s time interval, producing the required $\pi$ phase shift between states $\vert g\rangle$ and $\vert e\rangle$. The resonant interaction then resumes for the time $T_2$. Figure~\ref{fig:revival}(b) shows (blue circles) the revival of the Rabi oscillations, for $T_1=13.4\ \mu$s, induced by the phase reversal. The contrast of the revival is slightly reduced by the experimental imperfections. These imperfections are well understood and measured, as shown by the agreement with the solid blue line resulting from a numerical model. 

The field amplitude $\beta$ to be measured is injected into \setup{C} at time $t=0$ during the time reversal phase flip, see Fig.~\ref{fig:setup}(d). Since at this time the atom is detuned from the cavity mode, it is quite impervious to the resonant injection. Full red circles in Fig.~\ref{fig:revival}(b) present the revival signal after the injection of an amplitude $\beta=~1$ and the solid red line corresponds to the numerical model. The phase shift between the red and blue curves is about $1.3\pi$ for $T_2=T_1$, in good agreement with our expectation: $1.23\pi$ for $D=1.94$ used here.


    \begin{figure}[t!]
    		\includegraphics[width=\columnwidth]{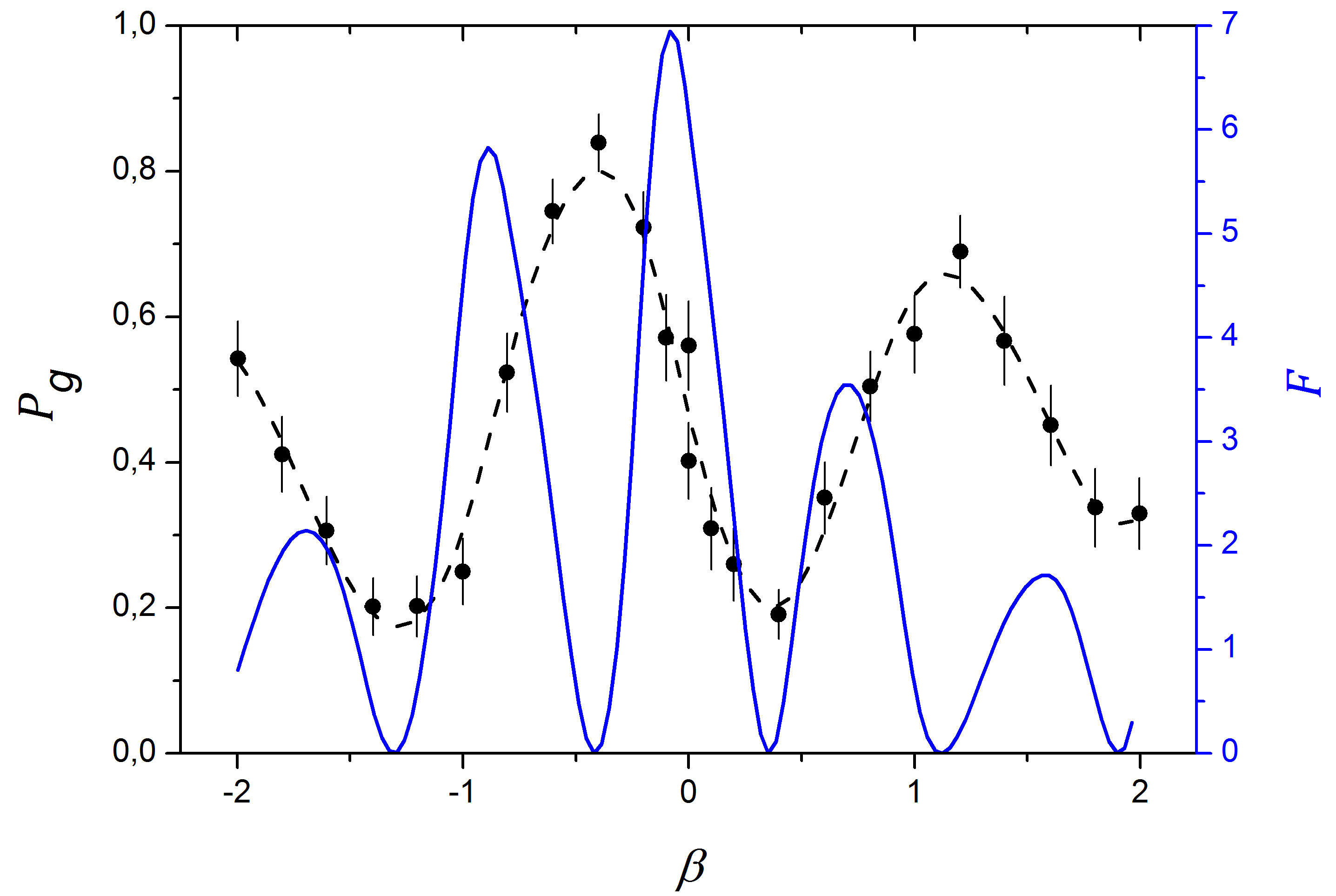}
  		\caption{Typical experimental interference signal $P_g(\beta)$. The points here are experimental for $T_1=12.0\,\mu$s and $T_2=13.5\,\mu$s. 
  		The dashed line is an interpolation from which we calculate the Fisher information (solid line related to the right $y$-axis).}
  		\label{fig:fringesExper}
	\end{figure}

\section{V. Results}

We have recorded, for fixed $T_1$ and $T_2$ values, the $P_g(\beta)$ signal as a function of the injected amplitude, making it possible to determine the available FI. For each $T_1$, we choose two $T_2$ values closest to $T_1$, such that $P_g=1/2$ for $\beta=0$. This mid-fringe condition provides the best sensitivity for the measurement of small displacements. 

The dots on Fig. \ref{fig:fringesExper} present the experimental signal as a function of $\beta$ for $T_1=12\,\mu$s and $T_2=13.5\,\mu$s. The dashed line is an interpolation with a polynomial function. From this continuous interpolation, we calculate the FI (solid line). It is, as expected, maximum for $\beta=0$ and reaches a value, 7, which is notably larger than $F_\textrm{SQL}=4$. The FI of the ideal signal of \eqref{eq:FisherFullyAnalytical2} would be~$14.9$. The information loss is due to dispersion in $T_1$ and $T_2$, that originates from the finite longitudinal spread of the atomic samples, having larger effect for larger displacements. 

  \begin{figure}[t]
    		\includegraphics[width=1.0\columnwidth]{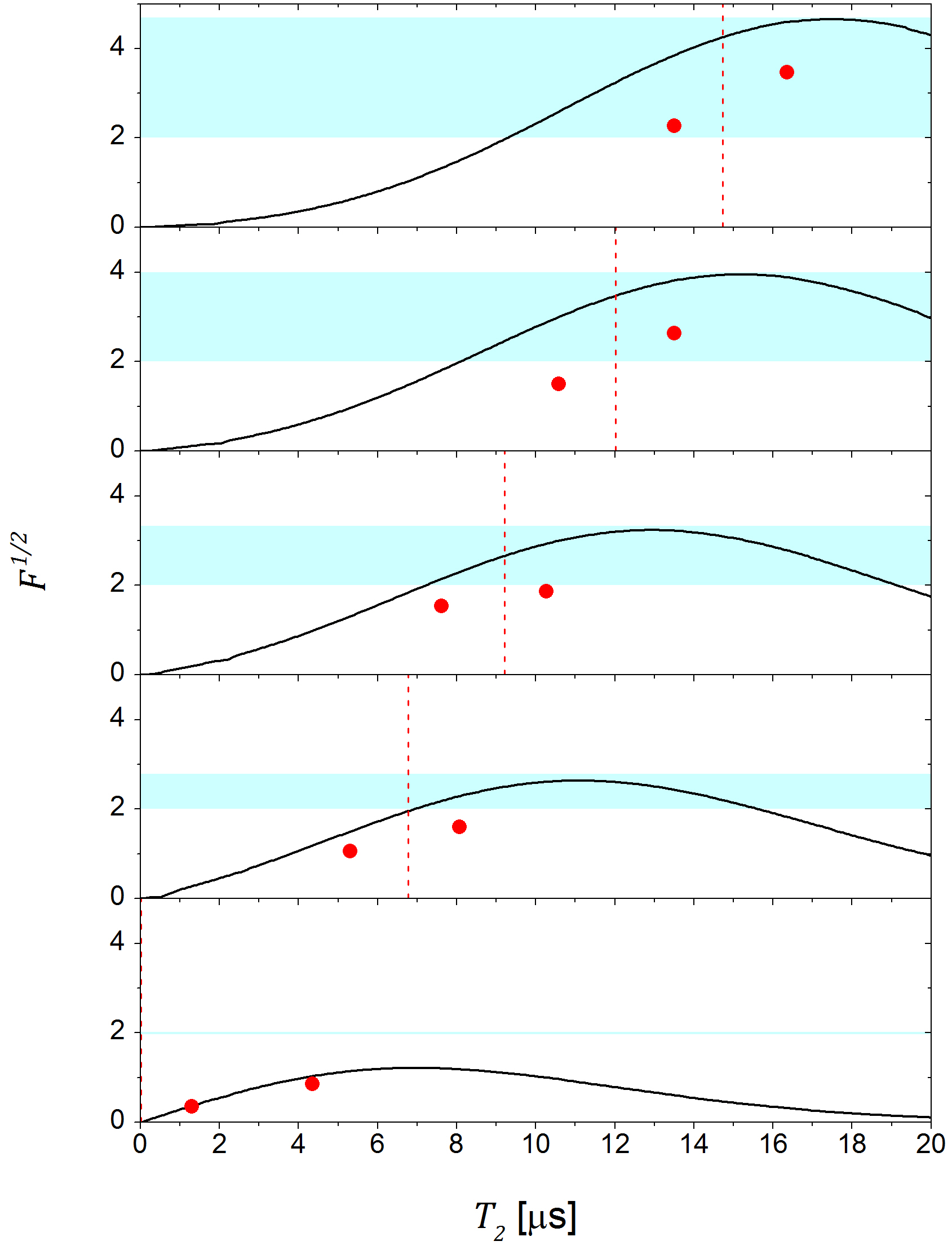}
  		\caption{Square-root Fisher information versus measurement time. Five subplots (from bottom to top) correspond to five values of the state preparation time $T_1$: 0~$\mu$s, 6.8~$\mu$s, 9.2~$\mu$s, 12.0~$\mu$s, and 14.7~$\mu$s, respectively. Circles are FI extracted from the measured data. Straight line is theoretical FI given by \eqref{eq:FisherFullyAnalytical2}. Horizontal bands correspond to the sub-Planck region of FI values between the SQL value of  $F_\textrm{SQL}=4$ and the QFI $F_Q$ given by \eqref{eq:QuantumFisher3}. Vertical dashed lines indicate the time of the complete revival ($T_2=T_1$) and are given for the reference.}
  		\label{fig:FisherData}
	\end{figure}
 
Figure~\ref{fig:FisherData} presents (red dots) the square root of the obtained FI (equal to $1/\Delta\beta^{(1)}$) as a function of $T_2$ for five values of $T_1$. As expected, for each $T_1$ the largest FI corresponds to the largest $T_2$ value. The solid curves are given by \eqref{eq:FisherFullyAnalytical2}. The horizontal bands correspond to the sub-Planck region with $F$ between $F_\textrm{SQL}=4$ and $F_Q=4+\Omega_0^2T_1^2$. As explained above, the theoretical FI is maximal for a value of $T_2$ larger than $T_1$. This maximum is very close to the QFI limit for all considered non-zero values of $D$. The convergence to optimality of this measurement process is thus quite fast.
 
The difference between the measured data and the theory is due to the spatial spread of the atomic samples leading to a dispersion in $\Omega_0$, $T_1$, and $T_2$.
This spread is increasingly disturbing when $T_2$ increases, preventing us from exploiting mid-fringe values of $T_2$ larger than those presented here.

Note that, for a coherent resource state ($T_1=0$), this measurement scheme is far from being optimal, since it provides a FI smaller than $F_\textrm{SQL}$ for all measurement durations $T_2$. In this simple case, the SQL can be straightforwardly obtained by a QND measurement of the photon number parity after the displacement starting from the vacuum state. It is easy to show that the FI of this measurement procedure equals exactly the QFI of the vacuum state. By increasing $T_1$, we enter into the non-classical regime and take benefit of the MQSS to overcome the SQL.

\begin{figure}[t!]
	\includegraphics[width=1.0\columnwidth]{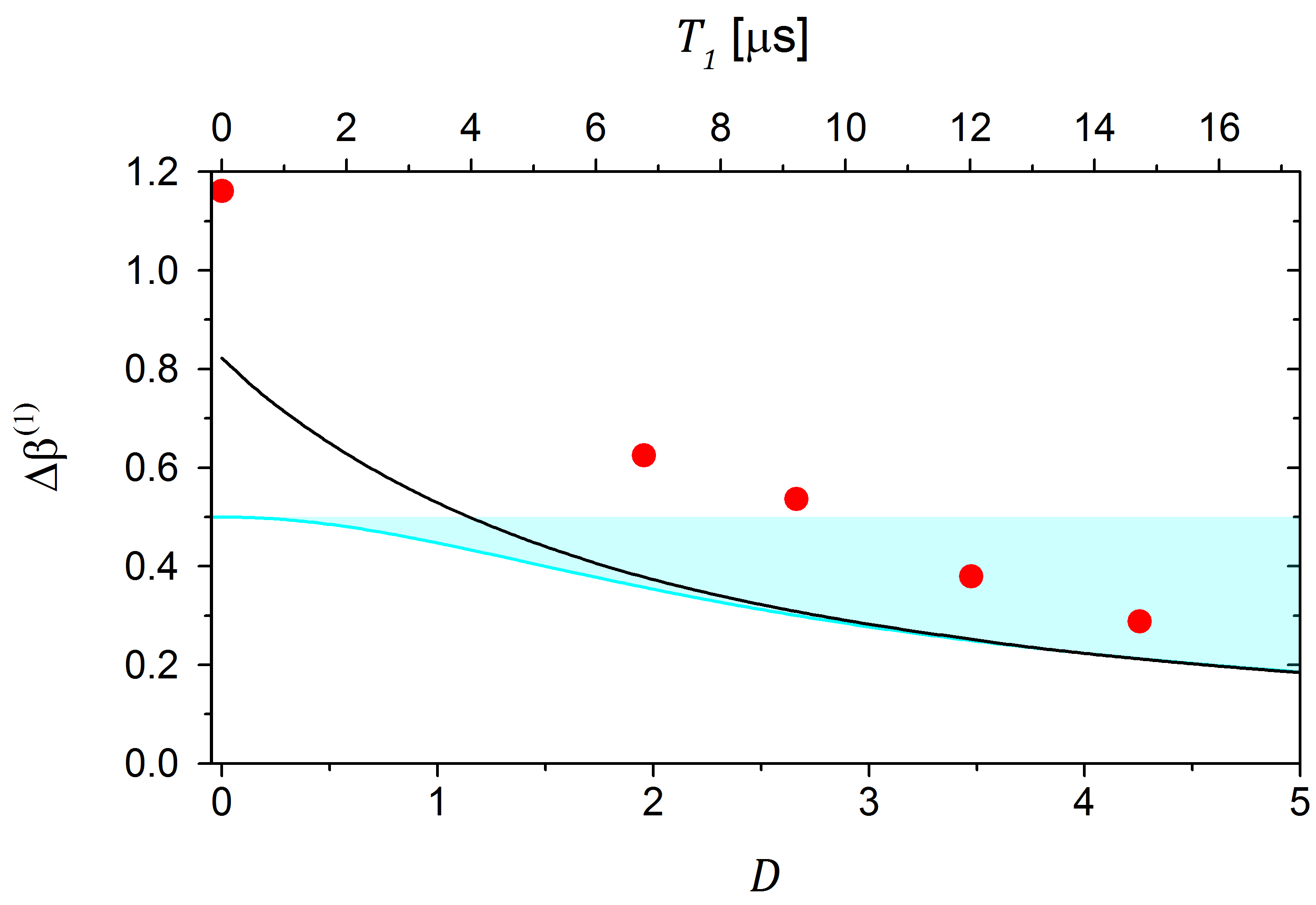}
  	\caption{Measurement precision versus preparation time and superposition size. Circles are $\Delta\beta^{(1)}$ extracted from the measured data. Straight line is $F$ of \eqref{eq:FisherFullyAnalytical2} maximized over measurement time $T_2$. Shaded (cyan) zone is the sub-Planck region bound from below by $\Delta\beta_{Q}={F_Q^{-0.5}}$ for a resource size $D$ and from above by $\Delta\beta_{SQL}=0.5$.}
  	\label{fig:FisherLimits}
\end{figure}

We summarize our main precision measurement results in Fig.~\ref{fig:FisherLimits}. We plot $\Delta\beta^{(1)}$ versus the preparation time $T_1$ and, equivalently, versus the resource MQSS size $D$. We choose for all $T_1$ values the largest $T_2$ in the pair. The red points are experimental. The solid line is the optimum theoretical FI  maximized over both $T_2$ and $\beta$. The blue band is the sub-Planck region, limited by the SQL from above and the QFI from below. The measurements with $D>3$ go beyond the SQL and approach the QFI for increasing $D$ values. 

\squeezetable
\begin{table}[h!]
  \centering
  \caption{\label{tab:Fisher}Fisher information for the largest state preparation time, $T_1=14.7\,\mu$s. The numbers to compare are bold: the ultimate upper bound set by $F_Q$ of the resource state, $F$ of the measured data, and $F_{SQL}$ of a coherent state setting the SQL bound.}
  \begin{ruledtabular}
\begin{tabular}{lcc}
    Fisher information & $T_2\!=\!13.5\mu$s & $16.3\,\mu$s\\
    \hline   
    $F_Q$ approximated by \eqref{eq:QuantumFisher3} 						 	 &  \multicolumn{2}{c}{21.6}\\
    $F_Q$ obtained from numerical integration 					 & \multicolumn{2}{c}{\textbf{20.5}}\\
    $F$ approximated by \eqref{eq:FisherFullyAnalytical2} & 13.9 & 21.0 \\
    $F$ from \eqref{eq:FisherFullyAnalytical2} with $5\%$ detection errors& 11.2 & 17.0 \\
    $F$ of the measured data & 5.1 & \textbf{12.0}\\
    $F_{SQL}$ ($F_Q$ of a coherent state)				 & \multicolumn{2}{c}{\textbf{4.0}}\\
  \end{tabular}
\end{ruledtabular}
\end{table}

For the measurement with the largest $D$ (\ie $T_1=14.7\,\mu$s), we give all the relevant theoretical and experimental values of $F$ and $F_Q$ in Table~\ref{tab:Fisher}. The first line corresponds to the prediction of the simple model of Section III. The second line takes into account in an explicit numerical simulation a small distortion of the coherent components during the resonant atom-field interaction, neglected in \eqref{eq:QuantumFisher3}. It reduces $F_Q$ by about $5\%$. The next lines  give three sets of FI values for the two $T_2$ values: 1~-~the ideal theoretical FI approximated by \eqref{eq:FisherFullyAnalytical2}; 2~-~the same FI with the detector imperfection taken into account; 3~-~the FI extracted from the measured data. The discrepancy between the expected and measured FI can be explained by the atomic sample spatial extension resulting in the non-negligible dispersion of experimental parameters in different experimental realisations. Even with all  these limitations, we obtain a measurement~$F$ three times higher than the SQL value. The corresponding improvement on the displacement measurement precision is $10\log (\sqrt{F/F_{SQL}}) \approx 2.4$~dB.


\section{VI. Conclusion}

We have presented an experimental scheme allowing to measure small field displacements with a precision exceeding the standard quantum limit. The scheme uses mesoscopic quantum superpositions generated and probed by the interaction of a single circular Rydberg atom with a field in a cavity. We analyze in detail the performance of the measurement in terms of Fisher information. The Fisher information carried by the measurement signal in principle approaches the quantum Fisher information of the initial resource state of the atom-cavity system. This shows that the measurement strategy is indeed optimal. Experimental imperfections to some extend reduce the observed Fisher information. However, it is still far above that of the standard quantum limit for the larger MQSS used. This experiment illustrates the potential of non-classical entangled states for quantum-enabled metrology.

The measurement precision is mainly limited by the available range of atom-cavity interaction times (total time limited to about 40 $\mu$s). We are setting up an experiment with slow Rydberg atoms in a cavity, which should allow us to reach much higher sensitivities, approaching the Heisenberg limit in this context. The principle of the measurement could also be transposed in the thriving circuit QED context, for instance, for the measurement of the amplitude of small propagating coherent fields.

\section{Acknowledgments}
\begin{acknowledgments}
The authors thank  W.~H.~Zurek for fruitful discussions. The authors acknowledge support from European Research Council (DECLIC project) and European Community (SIQS project). M.P.~acknowledges support from Direction G\'{e}n\'{e}rale de l'Armement (DGA) of the French
Ministry of Defence. S.G.~acknowledges support by the European Community FP7/2007-2013 Contract No.~626628
(Marie-Curie Actions). L.D.~acknowledges support from the Brazilian agencies CNPq, CAPES, FAPERJ, and the National Institute of Science and Technology for Quantum Information.
\end{acknowledgments}

\end{document}